\documentclass[aps,prd,preprint,superscriptaddress,amsmath,amssymb,showpacs]{revtex4-1}
\usepackage{dcolumn}
\usepackage{amsmath}
\usepackage{enumerate}
\usepackage{graphicx}
\usepackage{float}
\usepackage{physics}
\usepackage[colorlinks=true,allcolors=blue]{hyperref}
\begin{document}
	\title{Thermomagnetic Effects of Quark Matter in The NJL Model:
		Application of Regularization Schemes
	}
	
	\author{Xiang-Qiong Liu}
	\email{Corresponding author: }
	\affiliation{College of Mathematics and Physics, China Three Gorges University, Yichang 443002, China}
	
	\date{\today}

	\begin{abstract}
	In the SU(3) Nambu-Jona-Lasinio (NJL) model of a thermally magnetized medium, the regularization methods adopted for the thermodynamic potential and the mass gap equation are utilized to calculate the relevant thermodynamic quantities in thermomagnetic quark matter. When dealing with the thermodynamic quantities and the gap equation, three schemes can be chosen, namely magnetic field-independent regularization, soft cut-off regularization, and Pauli-Villars regularization. These three regularization schemes have different influences on the properties of thermomagnetic quark matter, and different choices of regularization schemes will also lead to different effects in the calculation of the properties of thermomagnetic quark matter.	
	\end{abstract}
	
	\maketitle
	
	\section{Introduction}\label{sec:01_intro}
	The Nambu-Jona-Lasinio (NJL) model is an effective nucleon-meson model constructed based on interacting Dirac chiral fermions \cite{Klevansky:1992qe,Buballa:2003qv,Witten:1984rs}. It still has a wide range of applications in the field of quantum chromodynamics (QCD). Especially in the low-energy state, since perturbative treatment methods cannot be applied, it is necessary to resort to effective models as substitutes. Thermomagnetic quark matter refers to the quark-gluon plasma phase that still exhibits the characteristics of perturbative interaction within a finite region beyond the hadron category, which is commonly known as quark matter. In special environments with extremely high temperatures and densities, such as the early universe and the interior of neutron stars, quarks and gluons can move freely, thus forming this unique form of matter \cite{Nambu:1960tm}\cite{Nambu:1961tp}.
	
	Under the condition of a strong magnetic field, the properties of quark matter will undergo extremely significant changes. Relevant studies have shown that a strong magnetic field can exert an influence on the chiral phase transition of quark matter, causing it to exhibit different dynamical masses and stability characteristics. With the help of the NJL model, in-depth investigations can be carried out on the dynamical masses and stabilities of two-flavor and three-flavor quark matter under a strong magnetic field. By adopting appropriate regularization schemes to describe magnetized quark matter, these studies are of great benefit to a deeper understanding of the role played by the magnetic field in the phase transition process of quark matter, and further provide solid theoretical support for explaining the actual existence forms of quark matter \cite{Menezes:2008qt,Menezes:2009uc,Fraga:2008qn}.
	
	In the calculations of the NJL model, ultraviolet divergence may occur. This is because in the calculations of quantum field theory, when integrating over momentum, in the region of large momentum (short distance), the integral often tends to infinity. If left unprocessed, these infinite results will cause the theory to lose its physical meaning. The regularization method is an effective means to deal with such ultraviolet divergence. It restricts the behavior of the integral in the large momentum region by introducing a cut-off parameter \cite{poggio1987computational} or a special functional form, so that the integral converges and finite results are obtained. For example, when using three-dimensional cut-off regularization, it truncates the part in the momentum space that is greater than a certain cut-off momentum, changing the integral range from the original one to another, thus avoiding the divergence caused by the large momentum region \cite{Leibbrandt:1975dj}\cite{cortez2005method}.
	
	The regularization method \cite{girosi1995regularization,Mao:2016fha,Vogl:1991qt,Gandhi:2021cod,Kunihiro:1987bb} helps us to deal with the contributions generated in the high-momentum (high-energy) region in a logical and reasonable manner within the framework of this effective theory. It enables us to, when focusing on low-energy physical phenomena, neglect the high-energy details that are relatively less crucial in the low-energy context by means of appropriate cut-off or adjustment parameters. In this way, we can more effectively explore a series of low-energy strong interaction phenomena such as chiral symmetry breaking and meson mass generation without getting deeply trapped in the complex physical processes of high energy.
	
	The structural layout of this article is as follows:  In Section \ref{sec:02 setup}, the theoretical framework of the three-flavor (NJL) model is presented in detail, and the formulas related to thermomagnetic quark matter such as thermodynamic quantities are derived.  Section \ref{sec:03 setup} mainly focuses on using three different regularization schemes to carry out targeted treatments on the NJL model. Section \ref{sec:04 summary} devotes efforts to numerical calculation work, using precisely quantified data to deeply explore and verify the characteristics and laws exhibited by the model under the action of different regularization schemes. Section \ref{sec:05 summary} comprehensively and systematically summarizes the relevant conclusions obtained from the various research works carried out in the previous parts.
	
	\section{Theoretical Part}\label{sec:02 setup}
	The Lagrangian of the SU(3)-NJL \cite{Klevansky:1992qe}\cite{Buballa:2003qv} model:
	\begin{equation}
		\begin{aligned}
			\mathcal{L}&=\overline{\psi}\left(-i{D}^{\mu }+\hat{M}\right)\psi+G\sum_{a=0}^{8}\left[\left(\overline{\psi}\lambda_{a}\psi\right)^{2}+\left(\overline{\psi}i\gamma_{5}\lambda_{a}\psi\right)^{2}\right]\\
			&-K\left[\mathrm{det}\overline{\psi}\left(1+\gamma_{5}\right)\psi+\mathrm{det}\overline{\psi}\left(1-\gamma_{5}\right)\psi\right]-\frac{1}{4}F^{\mu a}F_{\mu a}
		\end{aligned}
	\end{equation}
	where $\hat{M}=\text{diag}\left({{m}_{u}},~{{m}_{d}},~{{m}_{s}}\right)$ is the corresponding current quark mass matrix, $\psi ={{\left( {{\psi }_{u}},~{{\psi }_{d}},~{{\psi }_{s}} \right)}^{T}}$ represents the three-flavor quark field,  $G$, $K$ are the coupling constant. ${{\lambda }_{a}}=\sqrt{\frac{2}{3}}I,$ $I$ is the unit matrix, ${{\lambda }_{a}},\;a=1,2...8,$ represents the Gell-Mann matrices, $Q=\text{diag}\left( {{Q}_{u}},{{Q}_{d}},{{Q}_{s}} \right)$ is the charge matrix, with ${{Q}_{u}}=\frac{2}{3}e,~{{Q}_{d}}=-\frac{1}{3}e,~{{Q}_{s}}=-\frac{1}{3}e,$  $e$ is the proton charge. ${{D}^{\mu }}=\left( i{{\partial }^{\mu }}-Q{{A}^{\mu }} \right)$ is the covariant derivative. Consider the static and constant magnetic fields in three directions. Using the Landau gauge, we have ${{A}^{\mu }}={{\delta }_{\mu 2}}{{x}_{1}}B$. ${{A}^{\mu }}$ is the electromagnetic gauge field. ${{F}^{\mu \nu }}={{\partial }^{\mu }}{{A}^{\nu }}-{{\partial }^{\nu }}{{A}^{\mu }}$. \cite{Andersen:2014xxa,Miransky:2015ava,olver2010nist}
	
	Under the mean-field approximation:
	\begin{equation}
		\mathcal{L}^{\mathrm{MFA}}=\overline{\psi}\left(i{D}^{\mu }-\hat{M}\right)\psi-2\,G\,\left(\phi_{u}^{2}+\phi_{d}^{2}+\phi_{s}^{2}\right)+4\,K\,\phi_{u}\phi_{d}\phi_{s}
	\end{equation}
	where $\phi _{f}^{{}}$  is the expression of the condensate, the thermodynamic potential is 
	\begin{equation}
		\begin{aligned}
			\begin{aligned}\Omega=&-2Nc\sum_{f=u,d,s}\sum_{n=0}^{\infty}g_{n}\int\frac{p^{2}dp}{2\pi^{2}}\left[E_{f}+T\left(\ln\left(1+e^{-\beta(E_{f}+u_{f})}\right)+\ln\left(1+e^{-\beta(E_{f}-u_{f})}\right)\right)\right]\\
				&+2\,G\,\left(\phi_{u}^{2}+\phi_{d}^{2}+\phi_{s}^{2}\right)-4\,K\,\phi_{u}\phi_{d}\phi_{s}\end{aligned}
		\end{aligned}	
	\end{equation}
	${{E}_{f}}=\sqrt{{{p}^{2}}+{{M}_{f}}^{2}+2|{{q}_{f}}|Bn},~n=0,1,2...$, ${{g}_{n}}=2-{{\delta }_{n0}}$. ${{M}_{f}}$  is the mass of particle $f$. $\left| {{q}_{f}} \right|$ is the absolute value of the charge of particle $f$. ${{N}_{f}}=3$ and ${{N}_{c}}=3$, ${{u}_{f}}$ is chemical potential. In order to ensure that the thermodynamic potential in vacuum returns to zero, the normalized thermodynamic potential is defined as the effective potential.
	\begin{equation}
		\mathbf{\Omega}_{\mathrm{eff}}\left(T,\mu,M,B\right)=\Omega\left(T,\mu,M,B\right)-\Omega\left(0,0,M,B\right)
	\end{equation}
	 The gap equation is given by the following formula:
	\begin{equation}
		\begin{aligned}	
			{{M}_{u}}={{m}_{u}}-4\,G\,\phi _{u}^{{}}+2\,K\,\phi _{d}^{{}}\phi _{s}^{{}}  \\
			{{M}_{d}}={{m}_{d}}-4\,G\,\phi _{d}^{{}}+2\,K\,\phi _{s}^{{}}\phi _{u}^{{}}  \\
			{{M}_{s}}={{m}_{s}}-4\,G\,\phi _{s}^{{}}+2\,K\,\phi _{u}^{{}}\phi _{d}^{{}}  \\
		\end{aligned}
	\end{equation}
	The specific expression of the condensate is
	\begin{equation}
		\begin{aligned}	
			\phi_{f}=&-N_{c}\sum_{f=u,d,s}\left|q_{f}\right|B\sum_{n=0}^{\infty}g_{n}\int_{-\infty}^{\infty}\frac{dp}{\left(2\pi\right)^{2}}\frac{M_{f}}{\sqrt{p^{2}+M_{f}^{2}+2\beta_{f}n}}\\
			&+N_{c}\sum_{f=u,d,s}\left|q_{f}\right|B\sum_{n=0}^{\infty}g_{n}\int_{-\infty}^{\infty}\frac{dp}{\left(2\pi\right)^{2}}\frac{M_{f}\left(n\left(E_{n}\right)+\overline{n}\left(E_{n}\right)\right)}{\sqrt{p^{2}+M_{f}^{2}+2\beta_{f}n}}
		\end{aligned}
	\end{equation}
	where $\beta_{f}=|q_{f}|B$, $n\left(E\right)=\frac{1}{1+e^{\left(E_{f}-\mu_{f}\right)T}},
	~\overline{n}\left(E\right)=\frac{1}{1+e^{\left(E_{f}+\mu_{f}\right)T}}$. The crossover temperature or pseudo-critical temperature ${{T}_{pc}}$ at which chiral symmetry is partially restored is usually defined as the thermal sensitivity temperature ${{\chi }_{T}}$:
	\begin{equation}
		\chi_{T}=-m_{\pi}~\frac{\partial\sigma}{\partial T},~\sigma=\frac{<\overline{\psi}_{u}\psi_{u}>(B,T)~+ <\overline{\psi}_{d}\psi_{d}>(B,T)}{<\overline{\psi}_{u}\psi_{u}>(B,0)~+<\overline{\psi}_{d}\psi_{d}>(B,0)}
	\end{equation}
	Pressure $\mathbf{P}$ : According to ${{\text{ }\!\!\mathbf{\Omega}\!\!\text{ }}_{f}}=-\mathbf{P}_{f}={\varepsilon_{f}}-Ts-\underset{f}{\mathop \sum }\,{{\mu }_{f}}{{\rho }_{f}},$ there is
	\begin{equation}
		\mathbf{P}=\begin{pmatrix}P_u+P_d+P_s\end{pmatrix}-2\,G\begin{pmatrix}\phi_u^2+\phi_d^2+\phi_s^2\end{pmatrix}+4\,K\,\phi_u\phi_d\phi_s
	\end{equation}
	The normalized pressure is
	\begin{equation}
		\mathbf{P}_{\mathrm{eff}}\left(T,\mu,M,B\right)=\mathbf{P}\left(T,\mu,M,B\right)-\mathbf{P}\left(0,0,M,B\right)
	\end{equation}
	Energy density $\varepsilon $: It is obtained according to the formula
	\begin{equation}
		\begin{aligned}
			\varepsilon&=-T^{2}~\frac{\partial~(\mathbf{\Omega}_{eff}(T,u,M,B)/T)}{\partial~ T}|_{V}\\
			&=T~\frac{\partial~ (\mathbf{P}_{\mathrm{eff}}\left(T,\mu,M,B\right))}{\partial~ T}+\mu~\frac{\partial~ (\mathbf{P}_{\mathrm{eff}}\left(T,\mu,M,B\right))}{\partial~\mu}-\mathbf{P}_{\mathrm{eff}}\left(T,\mu,M,B\right)
		\end{aligned}
	\end{equation}
	Corresponding specific heat ${{C}_{V}}$:
	\begin{equation}
		C_{V}=\frac{\partial~\varepsilon}{\partial~ T}=-T~\frac{\partial^{2}~(\mathbf{\Omega}_{\mathrm{eff}}(T,u,M,B))}{\partial~ T^{2}}
	\end{equation}
	The square of the speed of sound $c_{s}^{2}$ under constant entropy $s$:
	\begin{equation}
		c_{s}^{2}=\frac{\partial~ \mathbf{P}}{\partial~\varepsilon}=\frac{\partial~(\mathbf{\Omega}_{eff}(T,u,M,B))}{\partial~ T}~/~T~\frac{\partial^{2}~(\mathbf{\Omega}_{eff}(T,u,M,B))}{\partial~ T^{2}}
	\end{equation}

	\section{Regularization Schemes}\label{sec:03 setup}
	 \begin{enumerate}[A.]
		\item Magnetic field-independent regularization (MFIR) \cite{Gandhi:2021cod}\cite{Allen:2015paa}: 
	\end{enumerate}
	Using the expression of the thermodynamic potential:
	\begin{equation}
		\Omega=\begin{pmatrix}\theta_u+\theta_d+\theta_s\end{pmatrix}+2G\begin{pmatrix}\phi_u^2+\phi_d^2+\phi_s^2\end{pmatrix}-4K\phi_u\phi_d\phi_s
	\end{equation}
	it is concluded that the thermodynamic potential \cite{Avancini:2020xqe} can be divided into three parts:
	\begin{equation}
		\theta_{f}^{\mathrm{vac}}=\frac{N_{c}}{8\pi^{2}}\left\{M_{f}^{4}\mathrm{ln}\left[\frac{\left(\Lambda+\epsilon_{\Lambda}\right)}{M_{f}}\right]-\epsilon_{\Lambda}\Lambda\left(\Lambda^{2}+\epsilon_{\Lambda}^{2}\right)\right\}
	\end{equation}
	\begin{equation}
		\theta_{f}^{\mathrm{mag}}=-\frac{N_{c}}{2\pi^{2}}(\left|q_{f}\right|B)^{2}\left[\zeta^{\prime}\left(-1,x_{f}\right)-\frac{1}{2}\left(x_{f}^{2}-x_{f}\right)\mathrm{ln}x_{f}+\frac{x_{f}^{2}}{4}\right]
	\end{equation}
	\begin{equation}
		\theta_{f}^{\mathrm{med}}=-\sum_{n=0}^{\infty}g_{n}N_{c}\frac{\left|q_{f}\right|B}{4\pi^{2}}\int_{-\infty}^{\infty}dp~T\left[\ln\left(1+e^{-(E_{f}+u_{f})/T}\right)+\ln\left(1+e^{-(E_{f}-u_{f})/T}\right)\right]
	\end{equation} 
	where ${{\epsilon }_{\text{ }\!\!\Lambda\!\!\text{ }}}=\sqrt{{{\text{ }\!\!\Lambda\!\!\text{ }}^{2}}+M_{f}^{2}}$, ~${{x}_{f}}=\frac{M_{f}^{2}}{2\left| {{q}_{f}} \right|B}$, $\zeta \left( z,q \right)$ is the Hurwitz-Riemann zeta function, ${\zeta }'\left( -1,{{x}_{f}} \right)=\frac{d\,\zeta \left( z,{{x}_{f}} \right)}{d\,z}{{|}_{z=-1}}.$
	The specific expression of the condensate is
	\begin{equation}
		\phi_{f}=\phi_{f}^{\mathrm{vac}}+\phi_{f}^{\mathrm{mag}}+\phi_{f}^{\mathrm{med}},~f=u,d,s.
	\end{equation}
	\begin{equation}
		\phi_{f}^{\mathrm{vac}}=-\frac{M_{f}N_{c}}{2\pi^{2}}\left[\Lambda\epsilon_{\Lambda}-M_{f}^{2}\mathrm{ln}\left(\frac{\Lambda+\epsilon_{\Lambda}}{M_{f}}\right)\right]
	\end{equation}
	\begin{equation}
		\phi_{f}^{\mathrm{mag}}=-\frac{M_{f}\left|q_{f}\right|BN_{c}}{2\pi^{2}}\left[\ln\Gamma\left(x_{f}\right)-\frac{1}{2}\ln\left(2\pi\right)+x_{f}-\frac{1}{2}\left(2x_{f}-1\right)\ln\left(x_{f}\right)\right]
	\end{equation}
	\begin{equation}
		\phi_{f}^{\mathrm{med}}=\sum_{n=0}^{\infty}g_{n}\frac{N_{c}\left|q_{f}\right|B}{4\pi^{2}}\int_{-\infty}^{\infty}dp~\frac{M_{f}}{E_{f}}\left(n(E)+n\left(\overline{E}\right)\right)
	\end{equation}
	where $\text{ }\!\!\Gamma\!\!\text{ } \left( {{x}_{f}} \right)$ is the Euler's totient function. The relevant calculation for the pressure \cite{Adel:2017uxi} is
	\begin{equation}
		P_{f}=P_{f}^{\mathrm{vac}}+P_{f}^{\mathrm{mag}}+P_{f}^{\mathrm{med}}
	\end{equation}
	\begin{equation}
		P_{f}^{\mathrm{vac}}=-\frac{N_{c}}{8\pi^{2}}\left\{M_{f}^{4}\mathrm{ln}\left[\frac{\left(\Lambda+\epsilon_{\Lambda}\right)}{M_{f}}\right]-\epsilon_{\Lambda}\Lambda\left(\Lambda^{2}+\epsilon_{\Lambda}^{2}\right)\right\}
	\end{equation}
	\begin{equation}
		P_{f}^{\mathrm{mag}}=\frac{N_{c}}{2\pi^{2}}(\left|q_{f}\right|B)^{2}\left[\zeta^{\prime}\left(-1,x_{f}\right)-\frac{1}{2}\left(x_{f}^{2}-x_{f}\right)\mathrm{ln}x_{f}+\frac{x_{f}^{2}}{4}\right]
	\end{equation}
	\begin{equation}
		P_{f}^{\mathrm{med}}=\sum_{n=0}^{\infty}g_{n}N_{c}\frac{\left|q_{f}\right|B}{4\pi^{2}}\int_{-\infty}^{\infty}dp~T\left[\ln\left(1+e^{-(E_{f}+u_{f})/T}\right)+\ln\left(1+e^{-(E_{f}-u_{f})/T}\right)\right]
	\end{equation}
	
	\begin{enumerate}[B.]
		\item Soft cut-off regularization 
	\end{enumerate}
	Soft cut-off regularization \cite{Chaudhuri:2022oru}\cite{fayazbakhsh2011phase} gradually suppresses the contributions from the high-momentum part through a smooth function. The functional form used is:
	\begin{equation}
		f_{\Lambda}\left(p\right)=\frac{1}{1+e^{\frac{\sqrt{p^{2}+2|q_{f}|Bn}-\Lambda}{0.05\Lambda}}}
	\end{equation}
	At the same time, for the thermodynamic potential and the temperature-related part of the condensate, both the regularized and non-regularized forms \cite{Xue:2021ldz} are used for processing. Define it as
	\begin{equation}
		\begin{aligned}
			\Omega&=-2Nc\sum_{f=u,d,s}\sum_{n=0}^{\infty}g_{n}\int\frac{p^{2}dp}{2\pi^{2}}\left[E_{f}+T\left(\ln\left(1+e^{-\beta(E_{f}+u_{f})}\right)+\ln\left(1+e^{-\beta(E_{f}-u_{f})}\right)\right)\right]\\
			&+2\,G\left(\phi_{u}^{2}+\phi_{d}^{2}+\phi_{s}^{2}\right)-4\,K\,\phi_{u}\phi_{d}\phi_{s}\\
			&=\Omega^{\mathrm{vac}}+\Omega^{T\mathrm{mag}}+2\,G\left(\phi_{u}^{2}+\phi_{d}^{2}+\phi_{s}^{2}\right)-4\,K\,\phi_{u}\phi_{d}\phi_{s}
		\end{aligned}
	\end{equation}
	\begin{equation}
		\begin{aligned}	
			\phi_{f}&=-N_{c}\sum_{f=u,d,s}\left|q_{f}\right|B\sum_{n=0}^{\infty}g_{n}\int_{-\infty}^{\infty}\frac{dp}{\left(2\pi\right)^{2}}\frac{M_{f}}{\sqrt{p^{2}+M_{f}^{2}+2\beta_{f}n}}\\
			&+N_{c}\sum_{f=u,d,s}\left|q_{f}\right|B\sum_{n=0}^{\infty}g_{n}\int_{-\infty}^{\infty}\frac{dp}{\left(2\pi\right)^{2}}\frac{M_{f}\left(n\left(E_{n}\right)+\overline{n}\left(E_{n}\right)\right)}{\sqrt{p^{2}+M_{f}^{2}+2\beta_{f}n}}\\
			&=\phi_{f}{}^{\mathrm{vac}}+\phi_{f}{}^{T\mathrm{mag}}
		\end{aligned}
	\end{equation}
	Then, using the regularized form
	\begin{equation}
		\begin{aligned}
		&\Omega=f_{\Lambda}\left(p\right)\Omega^{\mathrm{vac}}+f_{\Lambda}\left(p\right)\Omega^{T\mathrm{mag}}+2\,G\,\left(\phi_{u}^{2}+\phi_{d}^{2}+\phi_{s}^{2}\right)-4\,K\,\phi_{u}\phi_{d}\phi_{s}\\&\phi_{f}=f_{\Lambda}\left(p\right)\phi_{f}^{\mathrm{vac}}+f_{\Lambda}\left(p\right)\phi_{f}^{T\mathrm{mag}}
		\end{aligned}
	\end{equation}
	corresponding to the non-regularized form
	\begin{equation}
		\begin{aligned}
		&\Omega=f_{\Lambda}\left(p\right)\Omega^{\mathrm{vac}}+\Omega^{T\mathrm{mag}}+2\,G\,\left(\phi_{u}^{2}+\phi_{d}^{2}+\phi_{s}^{2}\right)-4\,K\,\phi_{u}\phi_{d}\phi_{s}\\&\phi_{f}=f_{\Lambda}\left(p\right)\phi_{f}^{\mathrm{vac}}+\phi_{f}^{T\mathrm{mag}}
		\end{aligned}
	\end{equation}
	
	\begin{enumerate}[C.]
		\item  Pauli-Villars regularization 
	\end{enumerate}
	In Pauli-Villas regularization, the functional form \cite{Pauli:1949zm} is used
	\begin{equation}
		f_{\mathrm{P.V}}\left(M_{f}\right)=\sum_{j=0}^{3}c_{j}f(\sqrt{M_{f}^{2}+j\Lambda^{2}}),\,c_{0}=1,\,c_{1}=-3,\,c_{2}=3,\,c_{3}=-1
	\end{equation}
	Then, using the regularized form
	\begin{equation}
		\begin{aligned}&\Omega=f_{\mathrm{P.V}}\left(M_{f}\right)\Omega^{\mathrm{vac}}+f_{\mathrm{P.V}}\left(M_{f}\right)\Omega^{T\mathrm{mag}}+2\,G\,\left(\phi_{u}^{2}+\phi_{d}^{2}+\phi_{s}^{2}\right)-4\,K\,\phi_{u}\phi_{d}\phi_{s}\\&\phi_{f}=f_{\mathrm{P.V}}\left(M_{f}\right)\phi_{f}^{\mathrm{vac}}+f_{\mathrm{P.V}}\left(M_{f}\right)\phi_{f}^{T\mathrm{mag}}
		\end{aligned}
	\end{equation}
	corresponding to the non-regularized form
	\begin{equation}
		\begin{aligned}&\Omega=f_{\mathrm{P.V}}\left(M_{f}\right)\Omega^{\mathrm{vac}}+\Omega^{T\mathrm{mag}}+2\,G\,\left(\phi_{u}^{2}+\phi_{d}^{2}+\phi_{s}^{2}\right)-4\,K\,\phi_{u}\phi_{d}\phi_{s}\\&\phi_{f}=f_{\mathrm{P.V}}\left(M_{f}\right)\phi_{f}^{\mathrm{vac}}+\phi_{f}^{T\mathrm{mag}}
		\end{aligned}
	\end{equation}

	\section{Numerical Results}\label{sec:04 summary}
	The MFIR and soft cut-off adopt parameters \cite{Hatsuda:1994pi} $\Lambda =631.4\,MeV$, ${{m}_{u}}={{m}_{d}}=5.5\,MeV,~{{m}_{s}}=135.7\,MeV.$ $G=\frac{1.835}{{{\Lambda }^{2}}},~K=\frac{9.29}{{{\Lambda }^{2}}},$ and the Pauli-Villas (P.V) cut-off \cite{Carignano:2019ivp} uses $\Lambda =781.2\,MeV$, ${{m}_{u}}={{m}_{d}}=10.3\,MeV,~{{m}_{s}}=236.9\,MeV.$ $G=\frac{4.90}{{{\Lambda }^{2}}},~K=\frac{129.8}{{{\Lambda }^{2}}},$ while the rest of the parameter is ${{m}_{\pi }}=135\,MeV$.
	\begin{enumerate}[(a)]
		\item  The quark mass 
	\end{enumerate}
	\begin{figure}[H]
		\centering
		\includegraphics[width=0.5\linewidth]{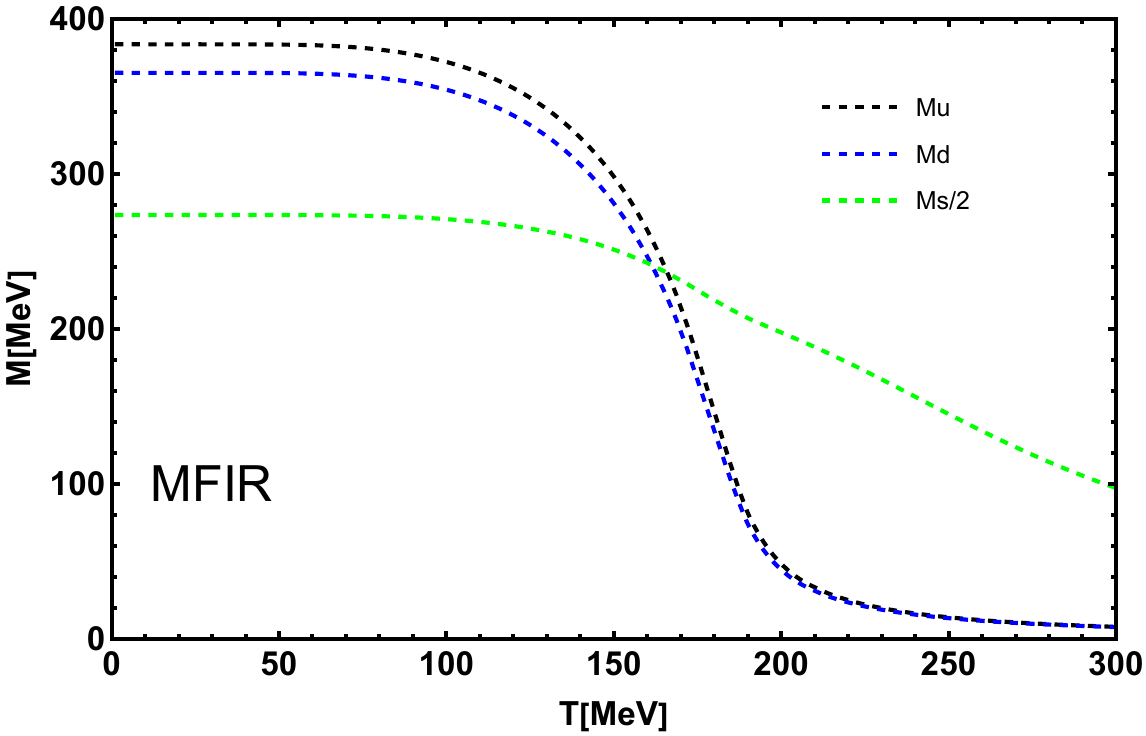}
		\caption{The relationship between M and T in magnetic-field-independent regularization}
		\label{fig:1}
	\end{figure}
	\begin{figure}[H]
		\centering
		\begin{minipage}{0.48\linewidth}
			\includegraphics[width=1.0\linewidth]{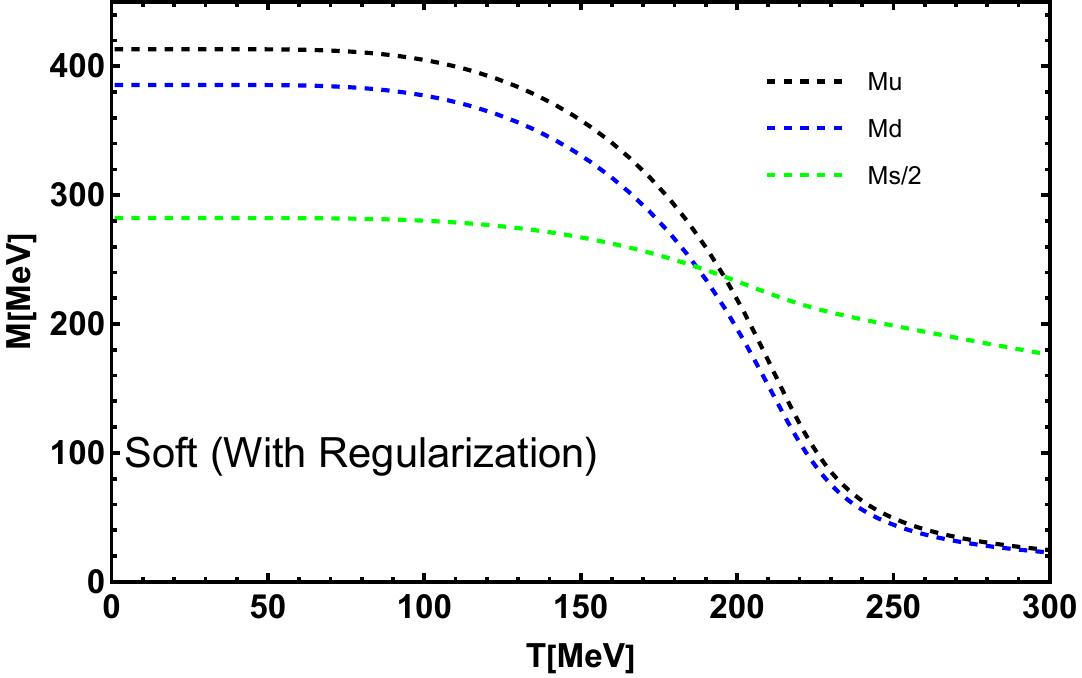}
		\end{minipage}\hfill
		\begin{minipage}{0.48\linewidth}
			\includegraphics[width=1.0\linewidth]{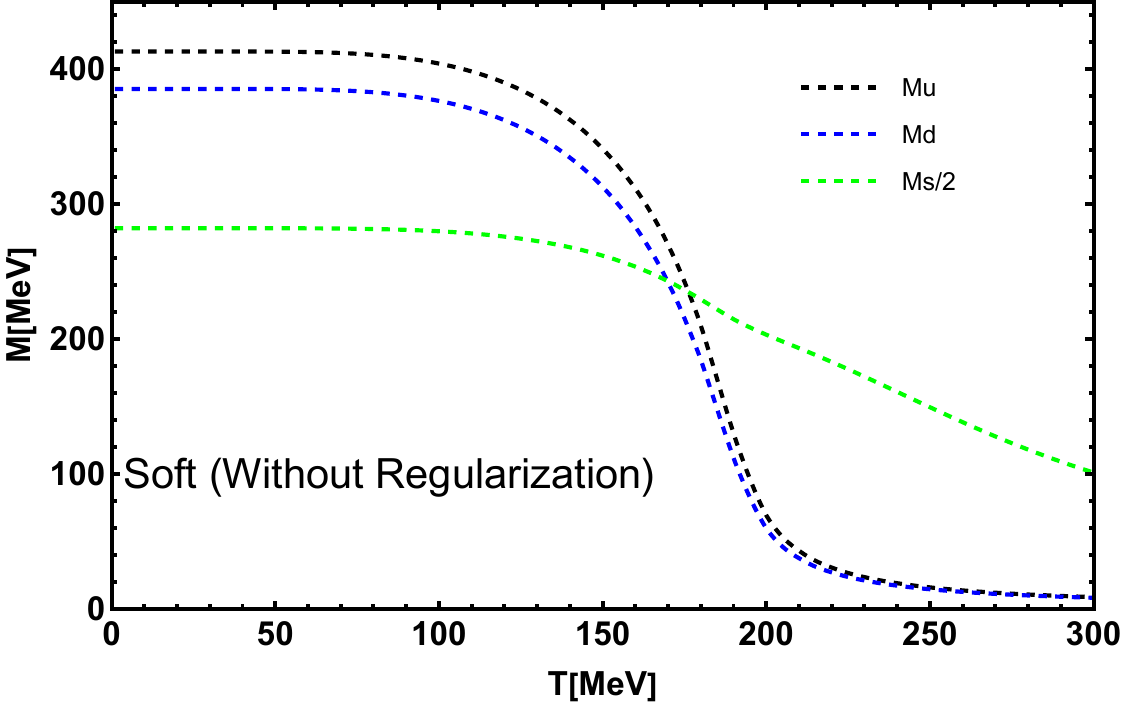}	
		\end{minipage}
		\caption{The relationship between M and T in the soft cutoff scheme: Left panel with regularization, right panel without regularization.}
		\label{fig:2}
	\end{figure}
	\begin{figure}[H]
		\centering
		\begin{minipage}{0.48\linewidth}
			\includegraphics[width=1.0\linewidth]{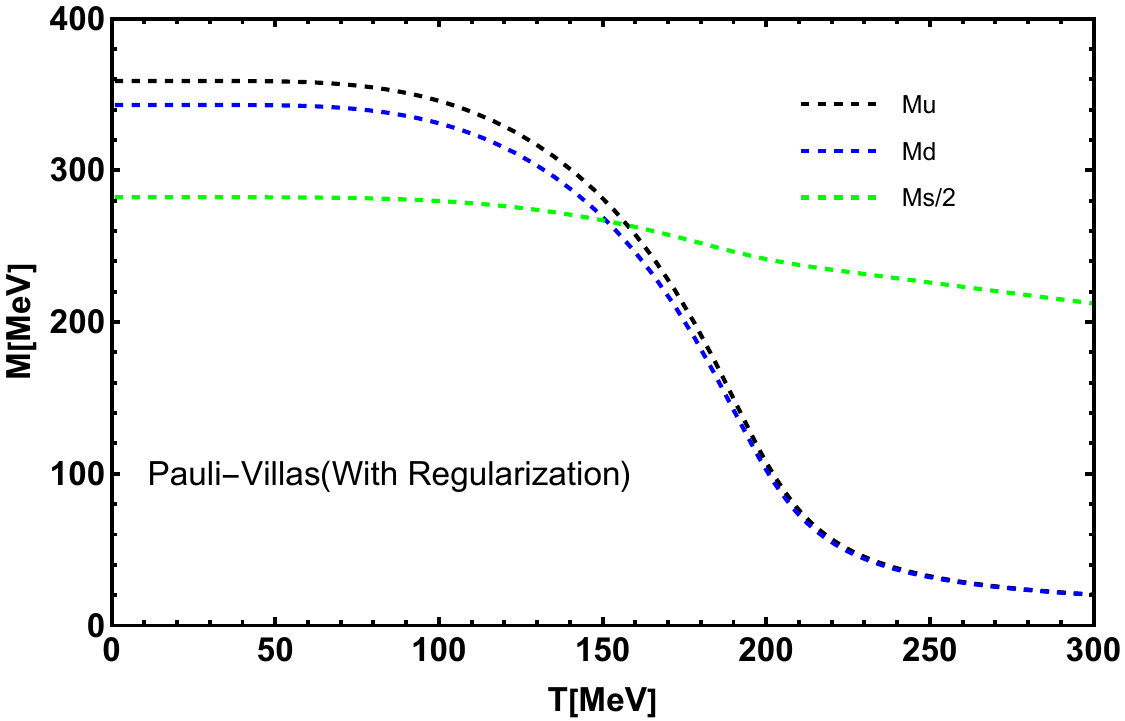}
		\end{minipage}\hfill
		\begin{minipage}{0.48\linewidth}
			\includegraphics[width=1.0\linewidth]{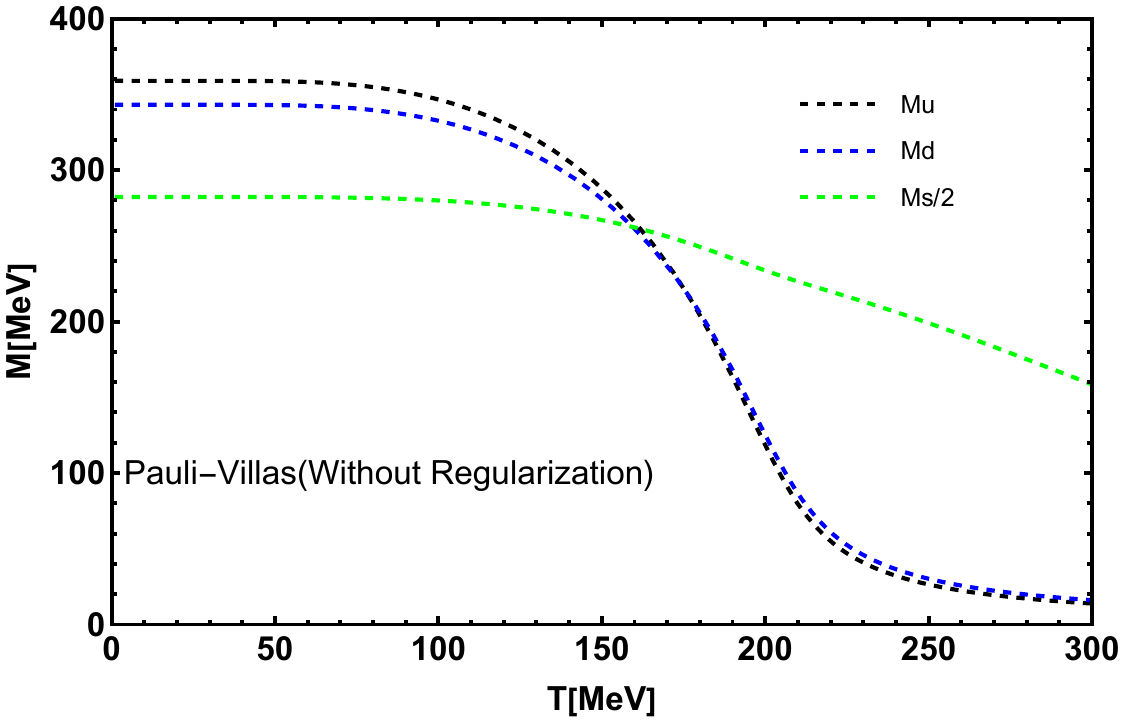}	
		\end{minipage}
		\caption{The relationship between M and T in the P.V cut-off scheme: Left panel with regularization, right panel without regularization.}
		\label{fig:3}
	\end{figure}
	 In Fig.\ref{fig:1}, Fig.\ref{fig:2}, Fig.\ref{fig:3}. in the quark mass-related diagrams obtained through calculation, regarding the $u$ quark, $d$ quark, and $s$ quark, their values are generally not much different numerically. Whether the soft cut-off and Pauli-Villas (P.V)  cut-off adopt the regularization scheme has almost no impact on the $u$ quark and $d$ quark, but has a greater impact on the $s$ quark. Obviously, the mass of the s quark decreases more at high temperatures. It can be concluded that the MFIR, soft cut-off, and P.V cut-off methods yield consistent results for the quark mass when no regularization scheme is applied to the temperature-dependent part. However, when a regularization scheme is applied to the temperature-dependent part, the primary and most significant effect is concentrated on the mass of the strange quark at high temperatures $T>200~MeV$.
	\begin{enumerate}[(b)]
		\item  For other related thermodynamic quantities 
	\end{enumerate}
	\begin{figure}[h]
		\centering
		\begin{minipage}{0.48\linewidth}
			\includegraphics[width=1.0\linewidth]{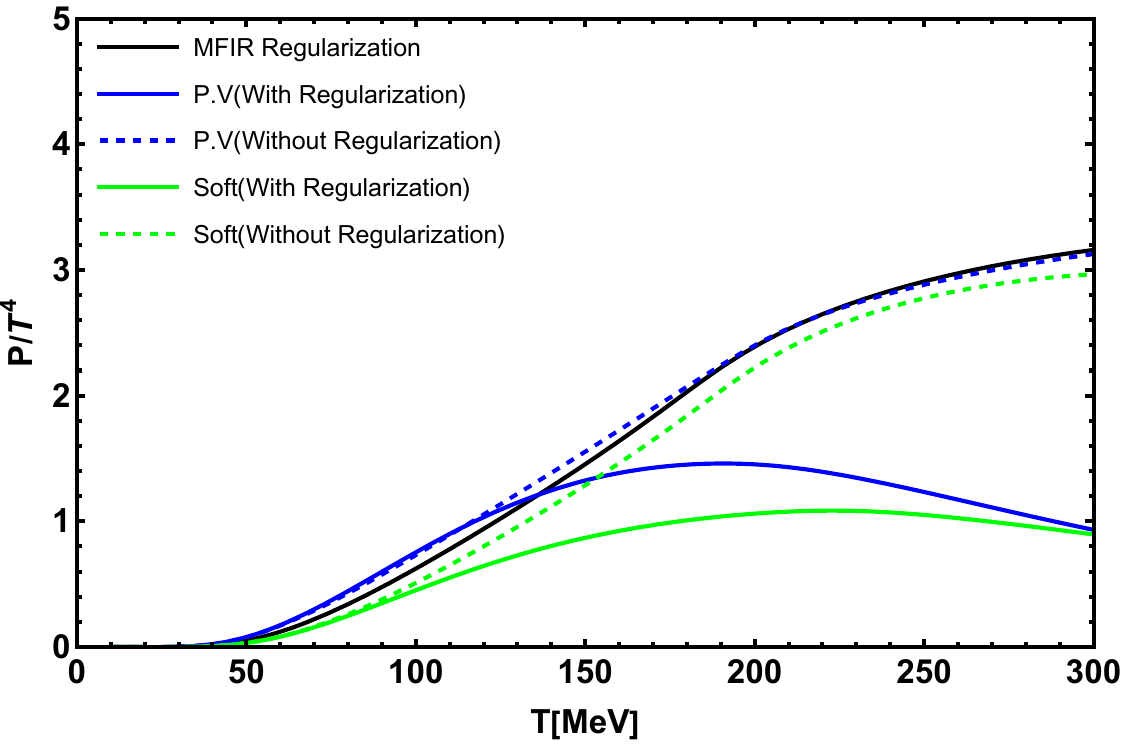}
		\end{minipage}\hfill
		\begin{minipage}{0.48\linewidth}
			\includegraphics[width=1.0\linewidth]{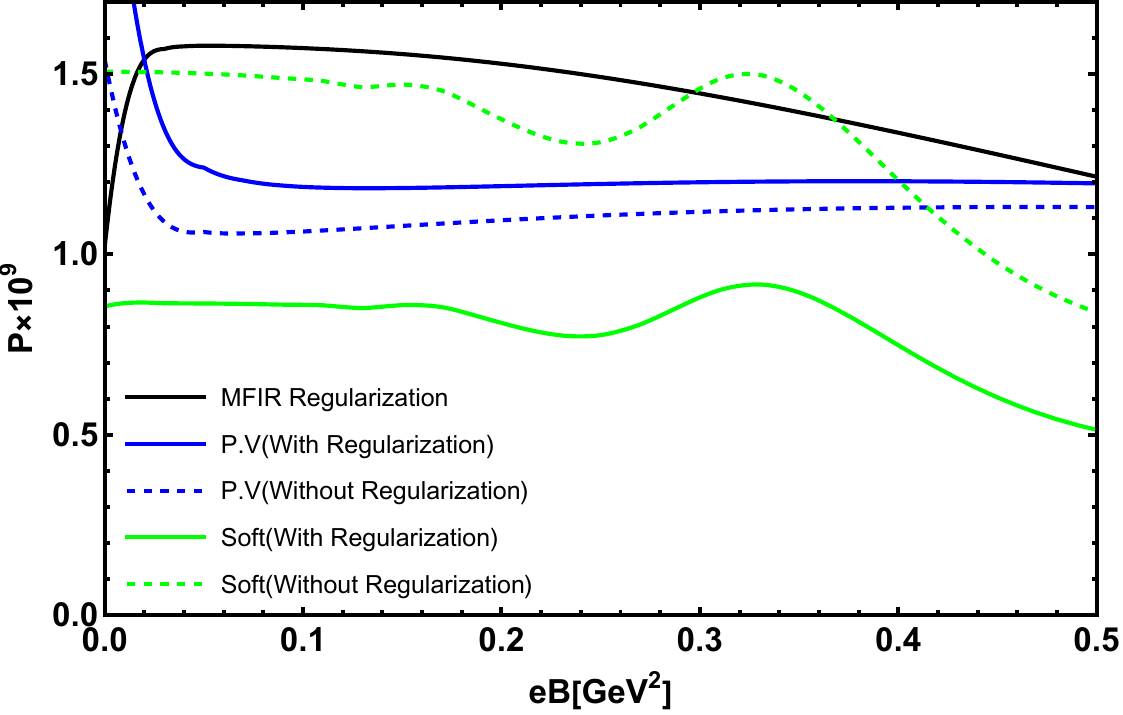}	
		\end{minipage}
		\caption{Left: The pressure varies with temperature. Right: The pressure varies with magnetic field.}
		\label{fig:4}
	\end{figure}
	As shown in Fig.\ref{fig:4}, in the relevant calculations regarding pressure with temperature, and magnetic field, it is observed that after the soft cut-off and P.V cut-off truncate the temperature-related parts, the pressure tends to decrease at high temperatures instead. This implies that regularizing the temperature-dependent part destabilizes the pressure, preventing it from equilibrating at high temperatures. In the graph showing the relationship between pressure and magnetic field, with the temperature fixed at $T=170~MeV$, it is found that under three different regularization methods, the changing trends of pressure are completely different. In the regularization method independent of the magnetic field, the pressure first shows an increasing trend and then decreases slowly. The P.V cut-off is just the opposite. In addition, under the soft cut-off, the pressure remains almost unchanged at low magnetic fields, but tends to $eB=0.2~Ge{{V}^{2}}$, and then decreases in a fluctuating manner. Regarding whether the regularization schemes are used for the soft cut-off and P.V cut-off and their impact on the pressure with respect to the magnetic field, it only affects the magnitude of the values but does not influence the overall trend change. Furthermore, the soft cut-off remains almost unchanged at low magnetic fields but shows oscillatory suppression as $eB=0.2~Ge{{V}^{2}}$, a feature attributed to vacuum-term regularization as demonstrated in \cite{Chaudhuri:2022oru}\cite{fayazbakhsh2011phase}. Whether regularization is applied to the temperature-dependent part in soft or P.V cut-off only affects the magnitude of the pressure under magnetic fields, not its general trend. The pressure-versus-field comparison confirms that different regularization schemes lead to substantial differences in the properties of hot magnetized quark matter.
	\begin{figure}[h]
		\centering
		\begin{minipage}{0.48\linewidth}
			\includegraphics[width=1.0\linewidth]{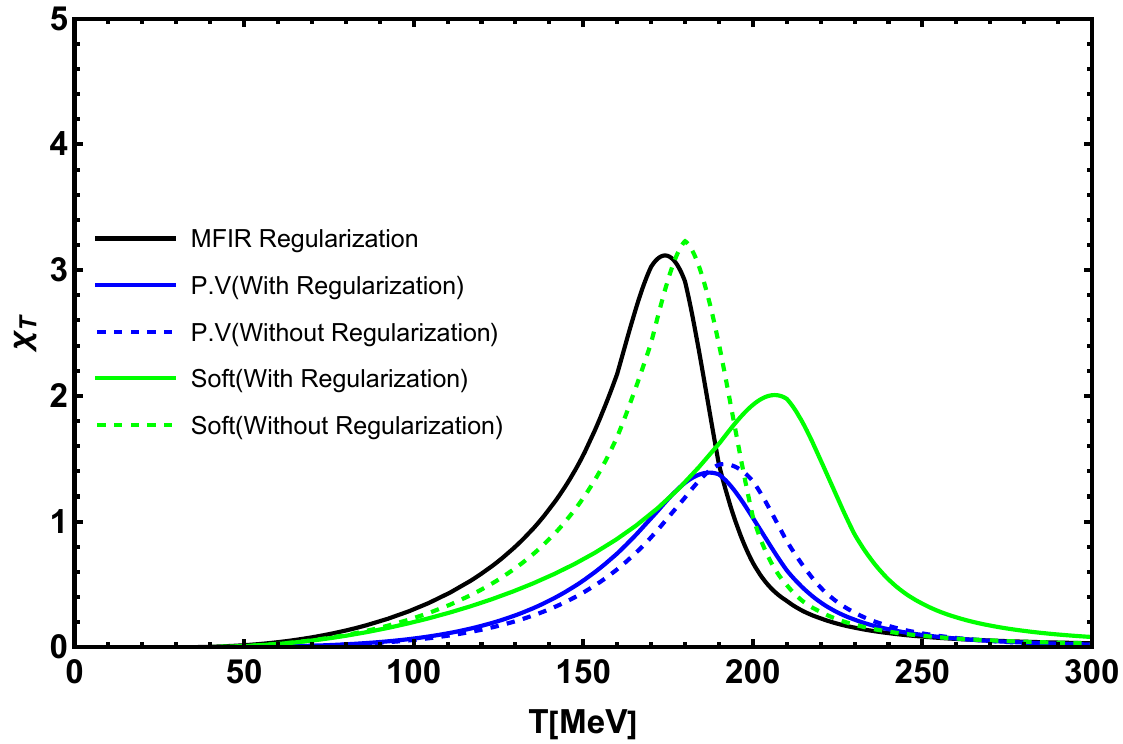}
			\caption{The variation of the thermal sensitivity temperature with temperature}
			\label{fig:5}
		\end{minipage}\hfill
		\begin{minipage}{0.48\linewidth}
			\includegraphics[width=1.0\linewidth]{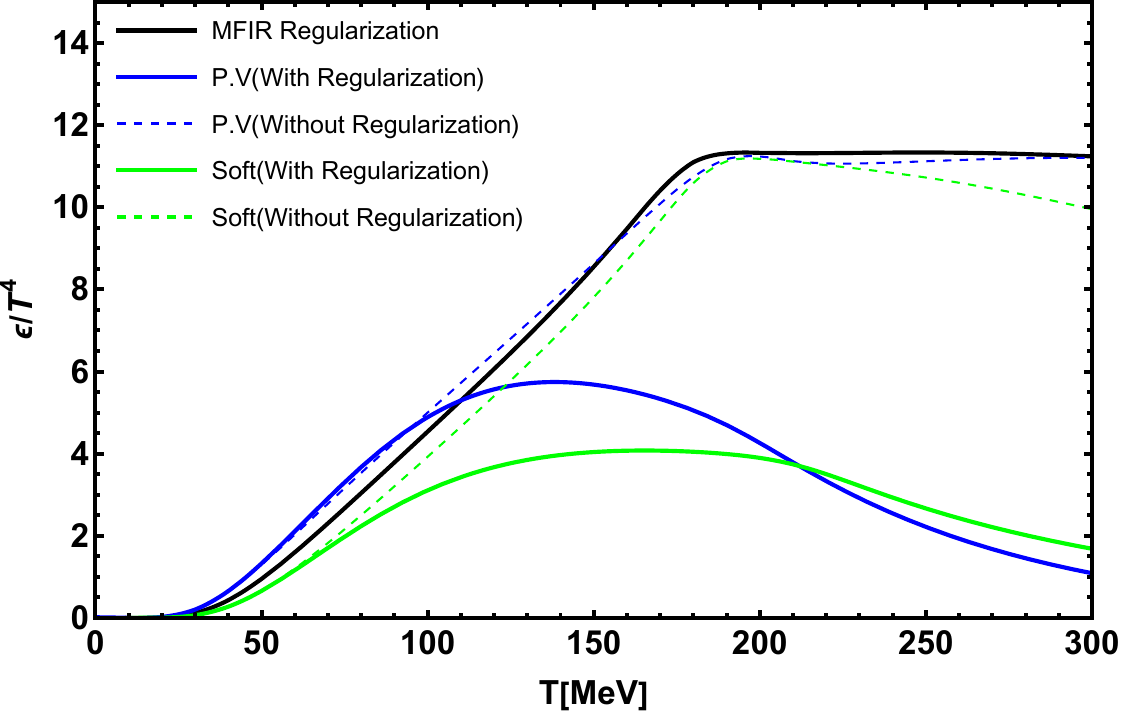}
			\caption{The variation of energy density with temperature}
			\label{fig:6}	
		\end{minipage}
	\end{figure}
	\begin{figure}[h]
		\centering
		\begin{minipage}{0.48\linewidth}
			\includegraphics[width=1.0\linewidth]{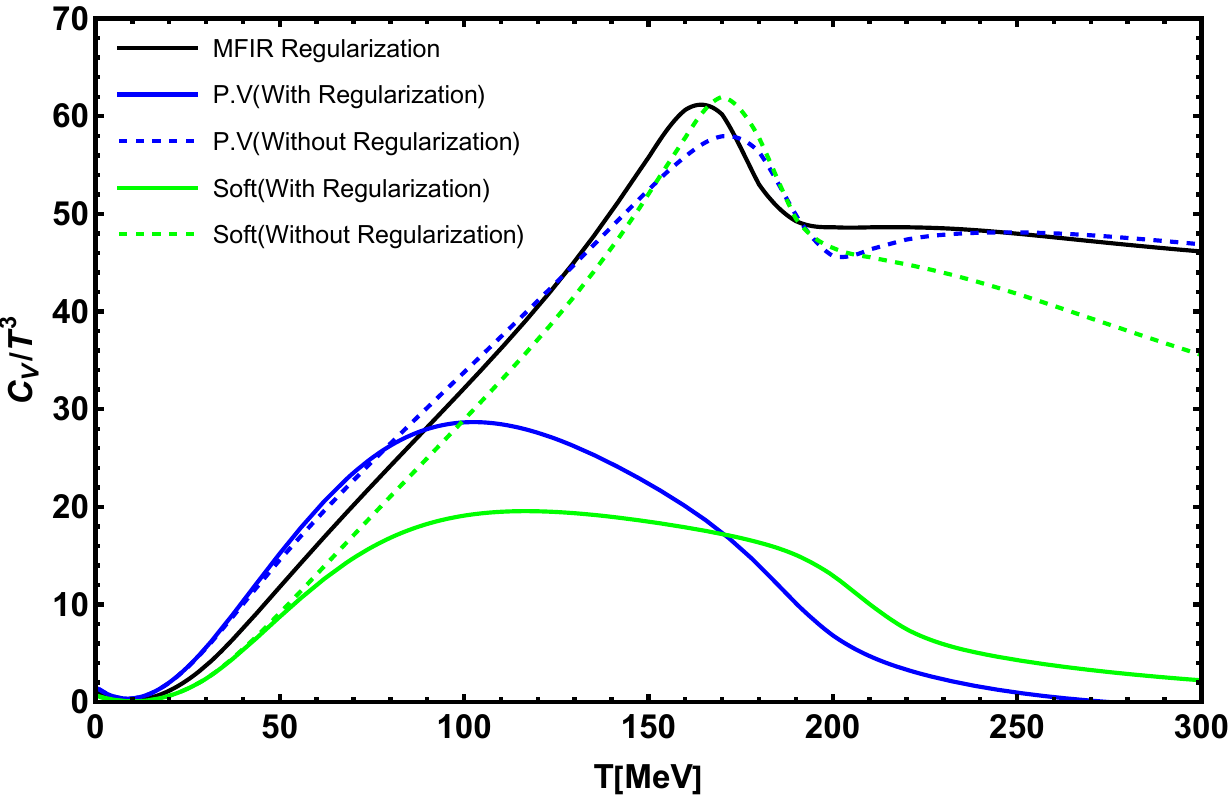}
			\caption{The variation of specific heat with temperature}
			\label{fig:7}
		\end{minipage}\hfill
		\begin{minipage}{0.48\linewidth}
			\includegraphics[width=1.0\linewidth]{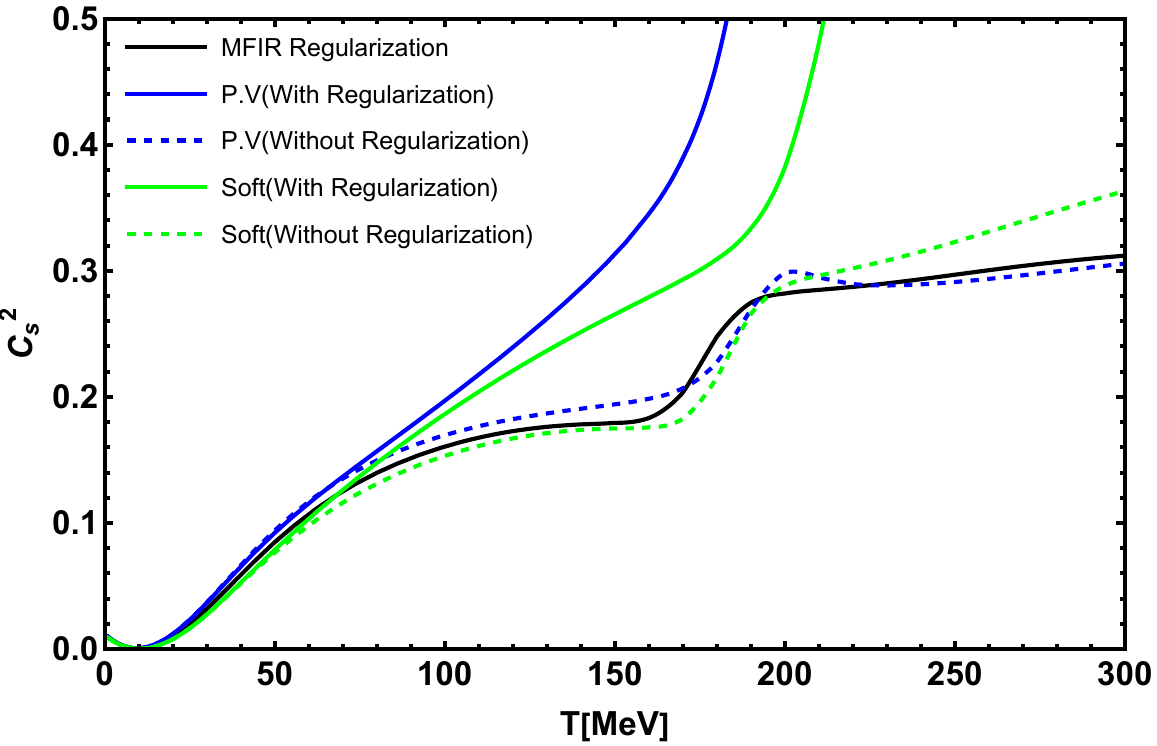}
			\caption{The variation of sound velocity with temperature}	
			\label{fig:8}
		\end{minipage}
	\end{figure}
	
	Fig.\ref{fig:5}, Fig.\ref{fig:6}, Fig.\ref{fig:7}, Fig.\ref{fig:8} corresponds to the thermodynamic quantities related to the phase transition temperature, energy, specific heat, and the speed of sound, etc. In the phase transition temperature plot in Fig.\ref{fig:5}, whether the P.V cut-off applies regularization to the temperature-dependent part has almost no effect on the peak position, whereas the soft truncation shows a noticeable shift in the peak depending on regularization. This suggests that the P.V cut-off 's impact on the phase transition temperature remains relatively stable, largely unaffected by whether regularization is applied to the temperature-dependent part.
	
	Using the MFIR results as the baseline, we compared the data obtained from soft cut-off and P.V cut-off without applying regularization to the temperature-dependent part. The analysis reveals that when computing thermodynamic quantities such as energy, specific heat, and speed of sound in Fig.\ref{fig:6}, Fig.\ref{fig:7}, Fig.\ref{fig:8}, the soft cut-off (without temperature regularization) exhibits noticeable deviations in the high-temperature regime $T>200~MeV$ compared to the other two regularization schemes. In contrast, the P.V cut-off (without temperature regularization) does not display such pronounced discrepancies. This observation highlights that omitting regularization in the temperature-dependent part under the soft cut-off scheme introduces significant variations in thermodynamic quantities, whereas the P.V cut-off remains more consistent. The findings suggest that the soft cut-off 's treatment of high-temperature effects is more sensitive to regularization choices, leading to stronger scheme-dependent differences in the results.
	
	Regarding the corresponding thermodynamic quantities in Fig.\ref{fig:6}, Fig.\ref{fig:7}, Fig.\ref{fig:8} under soft cut-off and P.V cut-off schemes, it can be clearly observed that the results without regularization of the temperature-dependent part are consistent with the MFIR results. In contrast, the results obtained using regularization schemes exhibit discrepancies. For instance, at high temperatures, the energy and specific heat in Fig.\ref{fig:6}, Fig.\ref{fig:7} no longer tend toward equilibrium but instead show a decreasing trend, while the speed of sound in Fig.\ref{fig:8} no longer approaches the Stefan-Boltzmann limit of $\frac{1 }{3}$ but instead displays a sharp increase. This indicates that in the calculation of such thermodynamic quantities, applying regularization to the temperature-dependent part can lead to violations of causality and affect the judgment of thermodynamic properties.
	
	\section{Conclusion}\label{sec:05 summary}
	Through specific calculations of the relevant properties of thermomagnetic quark matter, we have found that, from a numerical perspective, the regularization scheme independent of magnetic field is easier to handle. This, in turn, makes it possible for more complex numerical calculations and enables us to separately see the importance of contributions from the medium and external magnetic field in a more transparent manner, as the separation of magnetic and non-magnetic contributions is accomplished in a precise way. Meanwhile, the results obtained from the soft cut-off and Pauli-Villas regularization schemes also show an equivalent effect trend in dealing with the thermomagnetic effects of quark matter in the NJL model.
	
	Through a detailed comparative analysis of these data, it can be clearly recognized to a considerable extent that each of the three regularization schemes adopted has its own unique advantages and, at the same time, inevitably has certain limitations. In terms of calculating the relevant properties of thermomagnetic quark matter, the regularization scheme independent of magnetic field can separate the magnetic and non-magnetic contributions. Therefore, in the process of dealing with the NJL model, the data generated exhibits the most convenient and clear characteristics. However, when calculating the influence of other physical quantities on thermomagnetic quark matter, such as exploring the influence of the anomalous magnetic moment on the magnetic field, due to the precise separation characteristic of magnetic and non-magnetic contributions of this scheme, it is difficult to further carry out relevant calculation work on this basis. On the contrary, the Pauli-Villas cut-off without using the regularization scheme can smoothly promote relevant research in such a situation, although its computational complexity of data is significantly higher than that of other schemes. The soft cut-off regularization scheme, by virtue of its simple and efficient processing method, can effectively realize relevant calculations whether for the research of pure magnetic matter or considering the influence of other physical quantities. It should be noted, however, that since the magnetic field will have a fluctuating influence on related physical quantities such as pressure in a high-temperature environment, it leads to certain differences and discrepancies in the data obtained under high-temperature conditions.
	
	To sum up, there isn't a regularization scheme that can be universally applicable to all physical models. Different physical models vary in their own characteristics, application scopes and research objectives, and such differences further trigger diverse requirements for regularization schemes. The basic research on the three schemes, namely the regularization scheme independent of magnetic field, the soft cut-off regularization scheme and the Pauli-Villas cut-off without using the regularization scheme, can point out the direction for scholars dedicated to the research of the NJL model, helping them accurately judge which regularization scheme is more in line with their specific research directions and contents, so that they can make more reasonable and efficient choices in their research work and promote the in-depth development and breakthroughs in relevant fields.

	Data Availability Statement: No Data associated in the manuscript.

	\section*{References}
	
	\nocite{*}
	\bibliography{re}
	
\end{document}